\documentclass[prb,aps,twocolumn]{revtex4}
\usepackage{graphicx}
\usepackage{amsmath}

\begin{document}

\title{Impact of magnetic dopants on magnetic and topological phases \\
in magnetic topological insulators}
\author{Thanh-Mai Thi Tran and Duc-Anh Le}
\affiliation{Faculty of Physics, Hanoi National University of Education,  Hanoi, Vietnam}
\author{Tuan-Minh Pham, Kim-Thanh Thi Nguyen, and Minh-Tien Tran}
\affiliation{Graduate University of Science and Technology, Vietnam Academy of Science and Technology, Hanoi, Vietnam}
\affiliation{Institute of Physics, Vietnam Academy of Science and Technology,
Hanoi, Vietnam}

\begin{abstract}
A topological insulator doped with random magnetic impurities is studied. The system is modelled by the Kane-Mele model with a random spin exchange between conduction electrons and magnetic dopants. The dynamical mean field theory for disordered systems is used to investigate the electron dynamics. The magnetic long-range order and the topological invariant are calculated within the mean field theory. They reveal a rich phase diagram, where different magnetic long-range orders such as antiferromagnetic or ferromagnetic one can exist in the metallic or insulating phases, depending on electron and magnetic impurity fillings. It is found that insulator only occurs at electron half filling, quarter filling and when  electron filling is equal to magnetic impurity filling. However, non-trivial topology is observed only in half-filling antiferromagnetic insulator and quarter-filling ferromagnetic insulator. At electron half filling, the spin Hall conductance is quantized and it is robust against magnetic doping, while at electron quarter filling, magnetic dopants drive the ferromagnetic topological insulator to ferromagnetic metal. The quantum anomalous Hall effect is observed only at electron quarter filling and dense magnetic doping.
\end{abstract}

\maketitle

\section{Introduction}

Magnetic topological quantum materials are a relatively new type of materials, where topologically nontrivial electron properties coexist with magnetic ordering \cite{Burkov,Armitage,Yan,He,Weng}. These novel materials include magnetic topological insulators (MTIs), magnetic Weyl semimetals, magnetic Dirac semimetals... Experiments observed a remarkable quantization of the anomalous Hall conductance in a number of materials, for instance (Bi,Sb)$_2$(Se,Te)$_3$ doped with magnetic impurities \cite{He,Weng}.
In these materials the origin of the quantum anomalous Hall (QAH) effect relies  on the spin-orbital coupling (SOC) and magnetism \cite{YuFang,ZhangWang1,ChangXue,Jin,ZhangYao,Tien}.
Upon magnetic impurity doping, the spin exchange (SE) between the conduction electrons and magnetic moments induces a spontaneous magnetization at low temperature. The macroscopic magnetization could act on conduction electrons as the magnetic field in the anomalous Hall effect. The SOC keeps the topologically nontrivial band structure in the magnetic state. Without
the topologically nontrivial band structure, the magnetic impurity doping alone could not cause the QAH effect, for instance,  in the dilute magnetic semiconductors (DMSs), the magnetic dopants also induce a magnetic ordering, but the anomalous Hall conductance is not quantized, because the band structure of the DMSs is topologically trivial \cite{Jungwirth}.

Although the QAH effect emerges as a result of the interplay between the SOC and magnetism, its occurrence also depends on the filling level of conduction electrons and the concentration of magnetic dopants \cite{ZhangWang1,ChangXue,Jin,ZhangYao,Tien}.
The dopings of conduction electrons and magnetic impurities can drive both the topological and magnetic phase transitions \cite{ZhangWang1,ChangXue,Jin,ZhangYao,Tien}. In particular, first principle calculations showed successive topological and magnetic phase transitions from quantum spin Hall (QSH) state to QAH state, and then to ferromagnetic state, when the concentration of magnetic impurities increases \cite{Jin}. In these phase transitions magnetic dopants alter the SOC and the SE strengths, and the successive topological and magnetic phases are established as a result of the interplay between the SOC and the SE only \cite{Jin}.
Moreover, when magnetic impurities are doped, disorder and inhomogeneity are  inevitably  introduced.
As a consequence, the SE between conduction electrons and magnetic impurities is a random variable. Disorder of magnetic dopants can also induce random deviations of the magnetic moments from the macroscopic magnetization. Therefore, the induced magnetic ordering at low temperature depends on the concentration and the distribution of magnetic dopants. In many materials such as the DMSs or the colossal magnetoresistant materials, the doping of magnetic impurities is crucially important in determining the electronic and magnetic properties \cite{Timm,Dagotto}.
While the impact of electron and magnetic dopings on the magnetic and topological properties of the MTIs was experimentally studied, it has received less theoretical attention.

In this work, we study the impact of magnetic dopants on the magnetic and topological phases existing in the MTIs. We will construct a minimal model for MTIs. It should include at least two terms: the SOC which is responsible for the topologically nontrivial band structure and the SE between the magnetic dopants and conduction electrons that could induce a magnetic ordering at low temperature. The QAH effect emerges as a result of the interplay between the SOC and the SE \cite{He,Weng,Tien}.
In contrast to the first principle calculations, where the SOC and the SE are altered by magnetic dopants \cite{Jin}, in the minimal model the SOC and the SE are fixed upon magnetic doping. This allows us to study the direct impact of magnetic dopants on the the magnetic and topological properties.
In the previous study, a theoretical model for the MTIs was proposed \cite{Tien}. This model is based on a combination of the Kane-Mele model \cite{KaneMele1} and the SE between conduction electrons and magnetic impurities. It also looks like the double-exchange (DE) model with a SOC \cite{Dagotto,Zener}. The DE model is a minimal model proposed for different magnetic materials such as DMSs \cite{Jungwirth}, colossal magnetoresistance materials \cite{Dagotto,Zener}.
We found that the proposed model exhibits various magnetic insulating states, which occur at electron half and quarter (or three quarters) fillings, and they are topological insulator at appropriate values of the SE \cite{Tien}. However, the previous studies  assumed that the magnetic impurities are present at every lattice site \cite{He,Weng,Tien}. The doping of magnetic impurities away from the full filling and the disorder effect introduced by magnetic dopants were not previously considered. In this work, we study the impact of magnetic doping on the magnetic and topological properties, taking into account disorder and inhomogeneity introduced by magnetic dopants.
The dynamical mean field theory (DMFT) for disordered systems is used to study the proposed model \cite{Dobros1,Dobros2,Dobros3,Byczuk1,Byczuk2}.
Originally, the DMFT was introduced in order to correctly treat local electron correlations in infinite dimensional systems \cite{Metzner}. It has widely been used to study strong electron correlations \cite{GKKR}. Especially, the DMFT has successfully treated the SE in the DE-based models \cite{Furukawa,Yunoki,Kogan,Nham1,Tien1,Tien2,Nham2,Nham3}. By adopting the DMFT for disordered systems, we calculate
both the spontaneous magnetization and the topological invariant self-consistently. They reveal rich phase diagrams, depending on electron and magnetic dopings. It is found that the insulating state only occurs at electron (hole) half, quarter fillings, and at electron filling equaled to the concentration of magnetic dopants. However, the insulating state is topologically nontrivial only at electron (hole) half and quarter fillings.
At electron half filling, the QSH effect  is observed and it is robust against the magnetic impurity doping, while at electron quarter filling, the magnetic doping away from full magnetic filling suppresses the observed QAH effect. These findings reveal that magnetic dopants impact differently on the topological properties of the MTIs depending on electron filling.

The present paper is organized as follows. In Sec. II we describe the minimum model for MTIs and the DMFT for treating the SE and disorder introduced by magnetic dopants. The numerical results are presented in Sec. III. Finally, Sec. IV is the conclusion of the present work.

\section{Model and dynamical mean field theory}
\begin{figure}[b]
\includegraphics[width=0.3\textwidth]{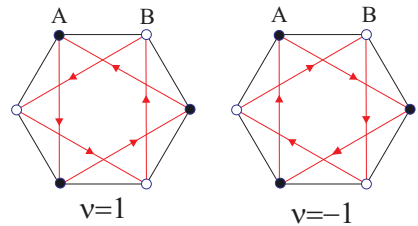}
\caption{(Color online) The sign structure $\nu_{ij}$ of the SOC term in the honeycomb lattice.}
\label{fig0}
\end{figure}

We consider a topological insulator doped with magnetic impurities. For the sake of simplicity, the topological insulator is modelled by the Kane-Mele Hamiltonian \cite{KaneMele1}. The Kane-Mele model consists of a nearest-neighbor hopping and an intrinsic SOC. In addition,  magnetic impurities are randomly distributed over the lattice. They are locally coupled with conduction electrons via a SE.
The Hamiltonian describing the model reads
\begin{eqnarray}
H &=&-t\sum\limits_{\langle i,j \rangle, \sigma }c_{i\sigma }^{\dagger }c_{j\sigma}
+i\lambda \sum\limits_{\left\langle \left\langle i,j\right\rangle\right\rangle ,s,s^{\prime }}
\nu _{ij}c_{is}^{\dagger }\sigma _{ss^{\prime}}^{z}c_{js^{\prime }} \nonumber \\
&&-\sum\limits_{i,ss^{\prime }} J_i \mathbf{S}_{i}c_{is}^{\dagger }
\boldsymbol{\sigma}_{ss^{\prime }} c_{is^{\prime }},
\label{ham}
\end{eqnarray}
where $c^{\dagger}_{i\sigma}$ ($c_{i\sigma}$) is the creation (annihilation) operator for electron with spin $\sigma$ at site $i$ of a honeycomb lattice. $\langle i,j \rangle$ and $\langle\langle i,j \rangle\rangle$ denote the nearest-neighbor and next-nearest-neighbor lattice sites, respectively. $t$ is the hopping parameter for the nearest-neighbor sites, and $\lambda$ is the strength of the intrinsic SOC. The sign $\nu_{ij}=\pm 1$ depends on the hopping direction, as shown in Fig. \ref{fig0}.
$\boldsymbol{\sigma}$ are the Pauli matrices. The honeycomb lattice is chosen, since the SOC in this lattice induces a topological insulating state \cite{KaneMele1}.
$\mathbf{S}_{i}$ is spin of magnetic impurity at lattice site $i$. We also treat it classically, as widely used in the studies of materials doped with magnetic impurities \cite{Dagotto,Furukawa,Yunoki,Kogan,Nham1,Tien1,Tien2,Nham2,Nham3}. Indeed, the magnetic moment of magnetic dopants is often big, for instance doped Mn ions in topological insulator Bi$_{2-x}$Mn$_{x}$Te$_3$ have the magnetic moment $\sim 4 \mu_{B}$ \cite{Cava}.  This consideration excludes any possibility of the Kondo effect \cite{Feng,Zhong,Tien3,Tien4}.
In fact, no any signature of the Kondo effect was observed in the MTIs.  $J_i$ is the strength of SE at lattice site $i$. We consider only the substitutional doping of magnetic impurities, and avoid any interstitial one. Indeed, first principle calculations show the substitutional doping is energetically more favorable than the interstitial one \cite{ZhangYao1}.
In contrast to the previous study \cite{He,Weng,Tien}, in this study magnetic impurities are randomly doped, and the SE is valid only on the lattice sites, where magnetic impurities are located. We  consider a binary distribution of magnetic dopants
\begin{eqnarray}
P(J_i) = (1-x)\delta(J_i) + x \delta(J_i-J) ,
\end{eqnarray}
where $x$ is the concentration of magnetic dopants.
Basically, only $x$ fraction of lattice sites has the local SE between conduction electrons and magnetic dopants. The parameter $x$ can also be interpreted as a disorder measurement of magnetic impurities.
However, both $x=0$ and $x=1$ correspond to the non-disordered cases.
When the magnetic impurities are absent ($x=0$), the proposed model returns to the Kane-Mele model \cite{KaneMele1}. The SOC causes a band gap at half filling, and the insulating state has an integer spin Chern number \cite{KaneMele1}. This yields the QSH effect.
In the opposite limit, $x=1$, magnetic impurities are present at every lattice site. Hamiltonian in Eq. (\ref{ham}) essentially describes the interplay between the SOC and the SE \cite{Tien}. It exhibits a coexistence of the QSH effect and antiferromagnetism at electron half filling and a ferromagnetic topological insulator at electron (hole) quarter filling \cite{Tien}. Between these two limiting cases, $0 < x <1$, magnetic dopants are randomly distributed, and they may drive the magnetic and topological phase transitions.
For classical impurity spin, the sign of the SE is irrelevant. Without loss of generality, we adopt the ferromagnetic sign $J>0$.

We divide the honeycomb lattice into two penetrating sublattices $A$ and $B$, as shown in Fig. \ref{fig0}. Then we denote $a_{I\sigma}$ ($b_{I\sigma}$) being the annihilation operator of electron at unit cell $I$ for the sublattice $A$ ($B$). We introduce a four-dimensional spinor
\begin{equation*}
\Psi_{I}=\left(
\begin{array}{c}
a_{I\uparrow } \\
b_{I\uparrow } \\
a_{I\downarrow } \\
b_{I\downarrow }%
\end{array}%
\right) .
\end{equation*}
For a fixed configuration of magnetic impurities, we introduce the Green function
\begin{eqnarray}
\mathbf{G}_{IJ}(i\omega_n,J_i) = -
\int\limits_{0}^{\beta} \!\! d \tau \; e^{- i \omega_n \tau} \langle {\mathcal{T}}
\Psi_{I} (\tau)  \Psi_{J}^{\dagger}\rangle,
\end{eqnarray}
where $\omega_n$ is the Matsubara frequency and $\beta=1/T$ is the inverse temperature.  Magnetic impurity disorder breaks the lattice translation invariance. However, the lattice translation invariance of the Green function is restored when Green-function averaging over the  magnetic impurity distribution is made. We obtain the averaged Green function in the momentum space
\begin{equation*}
\overline{\mathbf{G}}(\mathbf{k},i\omega_n)= \sum_{I,J} e^{-i \mathbf{k} \cdot (\mathbf{R}_I-\mathbf{R}_J)}
\overline{\mathbf{G}_{IJ}(i\omega_n,J_i)} ,
\end{equation*}
where the bar denotes the average over the magnetic impurity distribution.
The averaged Green function obeys the Dyson equation
\begin{equation*}
\overline{\mathbf{G}}(\mathbf{k},i\omega_n)= \left[ z-\mathbf{H}_0(\mathbf{k})-\boldsymbol{\Sigma }(\mathbf{k},i\omega_n)
\right] ^{-1} ,
\end{equation*}
where $\boldsymbol{\Sigma }(\mathbf{k},i\omega_n)$ is the self energy, and $\mathbf{H}_0(\mathbf{k})$ is the noninteracting and non-disordered Bloch Hamiltonian. The Bloch Hamiltonian reads
\begin{eqnarray}
\mathbf{H}_0(\mathbf{k}) = \left(
\begin{array}{cc}
\mathbf{h}_{\uparrow }(\mathbf{k}) & 0 \\
0 & \mathbf{h}_{\downarrow }(\mathbf{k})
\end{array}
\right),
\end{eqnarray}
where
\begin{eqnarray*}
\mathbf{h}_{\sigma }(\mathbf{k}) =\left(
\begin{array}{cc}
 \sigma \lambda \xi_{\mathbf{k}} & -t\gamma_{\mathbf{k}} \\
-t\gamma_{\mathbf{k}}^{\ast } & -\sigma \lambda \xi_{\mathbf{k}}%
\end{array}%
\right) ,
\end{eqnarray*}
and $\gamma_{\mathbf{k}}=\sum_{\delta} e^{i \mathbf{k} \cdot \mathbf{r}_{\delta}}$, $\xi_{\mathbf{k}}=i\sum_{\eta }\nu _{\eta }e^{i\mathbf{k}\cdot\mathbf{r}_{\eta }}$. Here $\delta$ and $\eta$ denote the nearest-neighbor and  next-nearest-neighbor sites of a given site in the honeycomb lattice, respectively. The self energy $\boldsymbol{\Sigma }(\mathbf{k},i\omega_n)$ includes all effects of interaction and disorder in an average manner. It renormalizes the dynamics of noninteracting and non-disordered conduction electrons.

We calculate the electron Green function by means of the DMFT. Here we will use the arithmetic average version of the DMFT for disordered systems \cite{Dobros1,Dobros2,Dobros3,Byczuk1,Byczuk2}. It basically is equivalent to the coherent potential approximation (CPA) \cite{GKKR}.
There is also a geometric average version of the DMFT that is usually called the typical medium theory \cite{Dobros1,Dobros2,Dobros3,Byczuk1,Byczuk2}.
The typical medium theory appropriately describes the Anderson localization in disordered systems.
In this work we focus on the effect of magnetic dopants on the magnetic and topological phases of MTIs,
where the Anderson localization is perhaps absent \cite{YuFang,ZhangWang1,ChangXue,Jin,ZhangYao}.
Apparently, the Anderson localization is induced by non-magnetic diagonal disorder or by off-diagonal disorder of conduction electron hopping \cite{Sheng1,Sheng2}. Such disorders are absent in the proposed Hamiltonian in Eq. (\ref{ham}).
Within the DMFT, the self energy depends only on frequency
$\boldsymbol{\Sigma }(\mathbf{k},i\omega_n) \rightarrow \boldsymbol{\Sigma }(i\omega_n)$.
The DMFT neglects nonlocal correlations at finite dimensions.
In the honeycomb lattice, the DMFT overestimates the critical value of the semimetal-insulator transition, but it is still capable to detect the insulating or magnetic states \cite{Sorella,Tran,WuLiu,Sorella1}. Due to the local nature, the DMFT  does not mix the different spin and sublattice sectors of the self energy, therefore $\boldsymbol{\Sigma }(i\omega_n)$ is a $4\times 4$ diagonal matrix.
The self energy obeys the Dyson equation
\begin{eqnarray}
\overline{G}_{a\sigma}(i\omega_n) = \mathcal{G}_{a\sigma}(i\omega_n) + \mathcal{G}_{a\sigma}(i\omega_n) \Sigma_{a\sigma}(i\omega_n) \overline{G}_{a\sigma}(i\omega_n) ,
\label{dyson}
\end{eqnarray}
where $a$ is the sublattice notation ($a=A, B$), and
$\overline{G}_{a\sigma}(i\omega_n)= \sum_{\mathbf{k}} \overline{G}_{a\sigma}(\mathbf{k},i\omega_n)/N$ is the local averaged Green function ($N$ is the number of sublattice sites).
The Green function $\mathcal{G}_{a\sigma}(i\omega_n)$ actually represents the effective dynamical mean field of conduction electrons. Within the DMFT, the self energy is determined from an effective single-site action, where $\mathcal{G}_{a\sigma}(i\omega_n)$ serves as the bare noninteracting Green function.
The action of the effective single site of sublattice $a$ with a fixed SE $J_a$ is
\begin{eqnarray}
\mathcal{S}_{a}(J_a)&=&-\sum\limits_{s}\int\limits_{0}^{\beta
}\int\limits_{0}^{\beta }d\tau d\tau ^{\prime }\Psi _{as}^{\dagger }(\tau ){%
\mathcal{G}}_{as}^{-1}(\tau -\tau ^{\prime })\Psi _{as}(\tau ^{\prime
}) \nonumber \\
&& -  \sum\limits_{\alpha ss^{\prime }}\int\limits_{0}^{\beta }d\tau J_{a} S^{\alpha
}(\tau )\Psi _{as}^{\dagger }(\tau )\sigma _{ss^{\prime }}^{\alpha }\Psi
_{as^{\prime }}(\tau ).
\label{action}
\end{eqnarray}
For classical impurity spin $\mathbf{S}$, this effective single-site action can exactly be solved \cite{Tien}.
Indeed, it is basically the action of an one-particle problem. Therefore, the DMFT is actually the CPA.
After solving the effective single-site problem, we obtain the local Green function
$G_{a\sigma}(i\omega_n,J_a)$ for a fixed SE $J_a$. The averaged local Green function can be calculated by
\begin{eqnarray}
\lefteqn{\overline{G}_{a\sigma}(i\omega_n)= \int d J_a P(J_a) G_{a\sigma}(i\omega_n,J_a)} \nonumber \\
&&= (1-x) G_{a\sigma}(i\omega_n,J_a=0) + x G_{a\sigma} (i\omega_n,J_a=J) .
\end{eqnarray}
Then, the self energy is determined by the Dyson equation (\ref{dyson}) again. So far, we have obtained the self consistent equations of the DMFT. They can be solved by simple iterations. After solving the DMFT equations, we obtain the self energy and the averaged Green function.
The spontaneous magnetizations of sublattice $A$ and $B$ are defined as
\begin{eqnarray*}
m_{A} = \frac{1}{2N} \sum_{I,\sigma} \sigma \overline{\langle a^\dagger_{I\sigma} a_{I\sigma} \rangle} , \\
m_{B} = \frac{1}{2N} \sum_{I,\sigma} \sigma \overline{\langle b^\dagger_{I\sigma} b_{I\sigma} \rangle} , \\
\end{eqnarray*}
where $\sigma=\pm 1$.  When $m_{A}= \pm m_{B} \neq 0$ the ground state is ferromagnetic or antiferromagnetic, respectively.
Due to the local nature, the DMFT does not mix different spin sectors of the Green function, hence the magnetization is not non-coplanar.
The topological property can be determined through the disorder-average transport \cite{Beenakker} or by the topological Bott index \cite{Loring,Loring1}.
The Bott index is defined in the real space with a realized configuration of random magnetic impurities. However, calculating the Bott index requires extensive numerical calculations. Instead of calculating the Bott index, here we will use the disorder-average approach proposed in Ref. \onlinecite{Beenakker}. Within this approach the self energy of the disorder-average Green function renormalizes the noninteracting and non-disordered Bloch Hamiltonian. Therefore, the renormalized Bloch Hamiltonian
$\mathbf{H}_{\text{eff}}(\mathbf{k})=\mathbf{H}_{0}(\mathbf{k})+\boldsymbol{\Sigma}(i0)=-[\overline{\mathbf{G}}(\mathbf{k},i0)]^{-1}$ determines the topological invariant, like in the non-disordered interacting case \cite{Wang}.
The topological invariant is determined by
\begin{eqnarray}
C_\nu =  \frac{1}{2\pi} \int d^2 k \mathcal{F}_{xy}^{\nu} , \label{chern}
\end{eqnarray}
where $\mathcal{F}_{ij}^{\nu}=\partial_{i} \mathcal{A}_{j}^{\nu} - \partial_{j} \mathcal{A}_{i}^{\nu}$,
$\mathcal{A}_{i}^{\nu}=i \langle \mathbf{k} \nu | \partial_{k_{i}} |\mathbf{k} \nu \rangle$, and  $|\mathbf{k} \nu \rangle$ is the orthonormalized eigenstate of matrix $\mathbf{H}_{\text{eff}}(\mathbf{k})$, corresponding to the eigenvalue $E_\nu(\mathbf{k})$.
This topological invariant is actually the Chern number of the effective Hamiltonian, where its renormalization is given by disorder and  interaction in the mean field approximation.
For weak disorder the disorder-average approach gives consistent results with the Bott index approach \cite{Guo1}.
In fact, the disorder-average approach has widely been used in determining the ground state topology \cite{Guo1,Guo,Song,Rotter,Refael}.
In numerical calculations one can use the efficient method of discretization of the Brillouin zone to calculate the Chern number in Eq. (\ref{chern}) \cite{Fukui}.

\section{Numerical results}

\begin{figure}[t]
\includegraphics[width=0.4\textwidth]{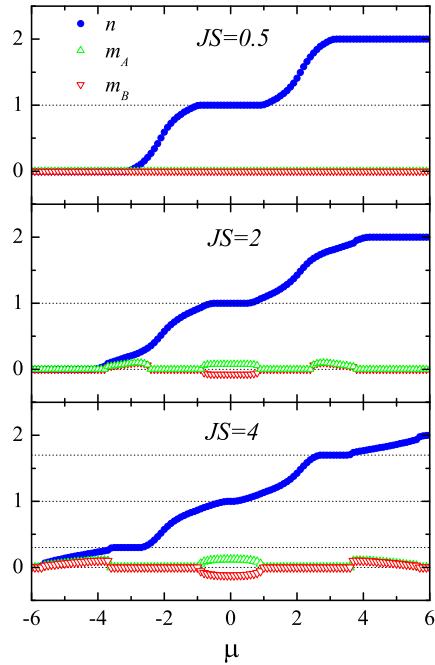}
\caption{(Color online) The electron filling $n$ and the sublattice magnetization $m_{A}$, $m_B$ via the chemical potential $\mu$ for different values of the SE at magnetic doping $x=0.3$ and SOC $\lambda=0.5$. For guiding the eye electron fillings $n=0.3, 1.0, 1.7$ are indicated by the horizontal dotted lines.}
\label{fig2}
\end{figure}

We numerically solve the DMFT equations by iteration for a given magnetic doping $x$. The numerical calculations are performed at fixed fictitious temperature $T=0.01$, which serves as the cell size of the Matsubara frequency mesh.
The emergence of magnetism and topology occurs in insulator, therefore we focus on detecting the insulating state. It is detected by a plateau in the curve $n(\mu)$, the dependence of electron filling on the chemical potential \cite{Tien}. Actually, the plateau reflects the band gap as well as the vanishing of the charge compressibility. They are the signals of the  insulating stability.

First, we consider the case of dilute magnetic doping ($x<0.5$).
In Fig.~\ref{fig2} we plot the dependence of electron filling $n$ and the sublattice magnetizations $m_A$, $m_B$ on the chemical potential $\mu$ for a small value $x$. It shows the plateau appearance at fillings $n=1$, $n=x$ and $n=2-x$ with appropriate values of the SE.
At electron half filling $n=1$, the system is transformed from an insulating state to a metallic state when the SE increases.
When the SE vanishes ($J=0$), the SOC opens a band gap in the electron structure \cite{KaneMele1}. A weak SE does not change the insulating state,
however, it reduces the band gap.  As a consequence, at an appropriate value of the SE, the band gap closes, and ground state becomes metallic.
At the same time, the SE also drives a magnetic phase transition \cite{Dagotto,Nham1}.
At electron half filling, it can induce an antiferromagnetic (AF) long range order with $m_A=-m_B \neq 0$ at low temperature \cite{Dagotto,Nham1}.
In a mean field picture, the AF magnetization can act on conduction electrons like a staggered magnetic field, and this field reduces the gap opened by the SOC. Indeed, when a staggered magnetic field is present, the energy spectra of conduction electrons become
\begin{eqnarray}
E(\mathbf{k}) = \pm \sqrt{t^2 |\gamma_{\mathbf{k}}|^2 + ( \lambda \xi_{\mathbf{k}} - h)^2},
\end{eqnarray}
where $h$ is the strength of the staggered magnetic field. Actually, in the mean field approximation $h \sim J$.
At the corners of the Brillouin zone $\mathbf{K}=2\pi(1/3,\pm 1/3\sqrt{3})$, $\gamma_{\mathbf{k}}$ vanishes, whereas $\xi_{\mathbf{k}}$ remains finite. When $h=\lambda \xi_{\mathbf{K}}$, the gap closes.
However, in contrast to the non-disordered magnetic case ($x=1$), at finite magnetic doping ($x<1$) strong SE does not open the band gap again, as can be seen in Fig. \ref{fig2}.
At strong SE, instead of antiferromagnetic insulator (AFI),
antiferromagnetic metal (AFM) is established. This is an
effect of magnetic impurity doping. Upon the magnetic doping, some lattice sites are free of the magnetic impurity occupation.
Therefore, at these sites conduction electrons are also free of the SE coupling. As a consequence, these conduction electrons give a contribution to the electrical conductivity.
However, this effect occurs only for strong SE, which aligns electron spins and magnetic moments in order to optimize the electron kinetic energy \cite{Zener}.
As can be seen in Fig.~\ref{fig2}, when the SE is strong, additional plateaus appear in the curve $n(\mu)$ at $n=x$ and $n=2-x$. Actually, $n=x$ and $n=2-x$ are equivalent due to the particle-hole symmetry. In this case, the concentrations of conduction electrons (holes) and of magnetic impurities are the same.
As we will see later, depending on magnetic doping $x$ and the SE strength, the ground state at $n=x$ (or $n=2-x$) may become magnetic.
In particular, at strong SE and dense magnetic doping ($x \gtrsim 0.8$), the ground state at $n=x$ and $n=2-x$ is AFI. In the limit $x \rightarrow 1$, these AFI states at $n=x$ and $n=2-x$ merge into the single AFI at electron half filling $n=1$.
As a consequence, at filling $n=x=1$, AFI occurs again when the SE is strong.
This can also be interpreted that the AFI at half filling in the full magnetic case ($x=1$) is actually split into two AFI states in the
electron and hole domains upon doping of magnetic impurities.

\begin{figure}[t]
\includegraphics[width=0.4\textwidth]{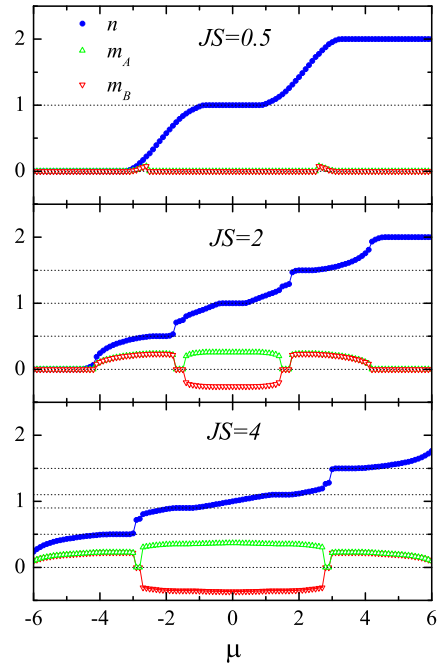}
\caption{(Color online) The electron filling $n$ and the sublattice magnetization $m_{A}$, $m_B$ via the chemical potential $\mu$ for different values of the SE at magnetic doping $x=0.9$ and SOC $\lambda=0.5$. For guiding the eye electron fillings $n=0.5, 0.9, 1, 1.1, 1.5$ are indicated by the horizontal dotted lines.}
\label{fig3}
\end{figure}

In dense magnetic doping ($x \gtrsim 0.8$),  additional plateaus in the curve of $n(\mu)$  are observed at $n=0.5$ and $n=1.5$ and strong SE, as can be seen in Fig. \ref{fig3}. Electron fillings $n=0.5$ and $n=1.5$ are equivalent due to the particle-hole symmetry. At electron quarter filling the insulating state is ferromagnetic because $m_A = m_B \neq 0$. This ferromagnetic insulator (FI) is also established in the non-disordered magnetic case ($x=1$) \cite{Tien}.
Figure \ref{fig3} also shows discontinuities of the electron filling $n$ and the sublattice magnetizations $m_A$, $m_B$ at certain values of the chemical potential. At these values of the chemical potential, the electron filling is uncertain, and actually the ground state is spontaneously  separated into two phases with the electron fillings corresponding to the extremes of the discontinuity in the curve $n(\mu)$.  This constitutes a phase separation \cite{Dagotto,Yunoki}.
The phase separation is not a disorder effect, because it also occurs in the non-disordered magnetic case $x=1$ \cite{Tien}. It occurs at the phase boundary between different symmetry phases, such as the magnetic and paramagnetic states.
In the magnetic state, the electron ground-state energy is optimized by aligning electron spins and magnetic moments through the SE coupling, while in the paramagnetic state electron spins are not aligned with the magnetic moments, and the optimization of the ground-state energy via the SE coupling is not operative \cite{Dagotto,Yunoki}. As a result of the competition of these two phases, a magnetic pattern is energetically formed at
the phase boundary. The phase separation often occurs in the DE model upon electron doping \cite{Dagotto,Yunoki}.

So far, we have observed the insulating state at electron (hole) fillings $n=1$, $n=0.5$, and $n=x$. However, the insulating state at quarter filling ($n=0.5$) occurs only when the doping of magnetic impurities closes to $x=1$. In the case of dilute magnetic doping, it is absent.
Electron fillings $n=1$ and $n=0.5$, where insulator is stable, reflect the number of occupied bands in the proposed model. At half filling $n=1$, the insulating state occurs when two lowest bands are fully occupied, whereas at quarter filling $n=0.5$, the full occupation of the lowest band yields the insulating state.
For other models, where the number of energy bands is larger, the filling condition for the insulating stability may be changed \cite{Bo}.

\subsection{Half filling $n=1$}

\begin{figure}[b]
\includegraphics[width=0.38\textwidth]{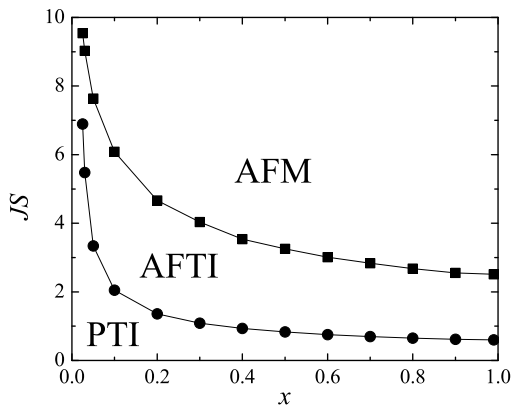}
\caption{Phase diagram at electron half filling $n=1$ ($\lambda=0.5$). Abbreviations PTI, AFTI, AFM denote paramagnetic topological insulator,  antiferromagnetic topological insulator, and antiferromagnetic metal, respectively. }
\label{fig4}
\end{figure}

In Fig. \ref{fig4}, we plot the phase diagram for a fixed SOC at electron half filling $n=1$. It shows that the insulating state exists regardless of magnetic impurity doping when SE is weak.
At weak SE the paramagnetic insulator (PI) is established. We have also calculated the Chern number defined in Eq. (\ref{chern}). It turns out this PI is topological with the spin Chern number $C=1$. Actually, it adiabatically connects to the $Z_2$ topological insulator in the non-interacting and non-disordered case $x=0$ \cite{KaneMele1}. The topological invariant is robust against the SE coupling  until the SE closes the band gap. On the other hand, the SE coupling aligns electron spins with the magnetic moments in order to optimize the ground state energy \cite{Dagotto,Zener}. When the SE strength is larger a certain
value, AF ordering is established at low temperature.
At electron half filling the ground state is insulating, hence there are no mediated itinerant electrons that can generate the magnetic long-range order by
the DE mechanism \cite{Dagotto,Zener}. However, the spontaneous magnetization in the insulating states can occur due to the direct coupling between the magnetic moments and electron spins through the van Vleck mechanism \cite{YuFang}.
We find that the AFI at half filling is also a topological insulator with the spin Chern number $C=1$.
Actually, the SE coupling drives only the magnetic phase transition from paramagnetic to antiferromagnetic state. Across this phase transition the topological invariant is not changed, because the band gap is still not closed by the SE.
The magnetic phase transition is quite similar to the one in the non-disordered magnetic case ($x=1$) \cite{Tien}.
With further increase of the SE coupling, the band gap is closed and the ground state is AFM, except for $x=1$, where the ground state is AFI.
As we have previously discussed, the AFI at strong SE in the non-disordered magnetic case ($x=1$) adiabatically connects to the merged insulating state at equal filling $n=x$ and $n=2-x$,
when $x \rightarrow 1$. Therefore, the ground state at magnetic dopings $x<1$ and $x=1$ has different origins.
Figure \ref{fig4} also shows that the non-trivial topology of the AF ground state at electron half filling is robust against magnetic doping. The  topological invariant remains the same regardless of magnetic doping $x$. This indicates that the QSH effect is protected even in the presence of magnetic dopant disorder as long as the band gap is still open.
Some MTI materials doped with magnetic impurities favor the AF state, for instance, first-principle calculations show an AF state in
Bi$_2$Se$_3$ doped with Fe ions \cite{ZhangYao1}. However, it is still challenge to find the coexistence of the QSH effect and AF ordering in MTIs.

\subsection{Quarter filling $n=0.5$}

\begin{figure}[b]
\includegraphics[width=0.48\textwidth]{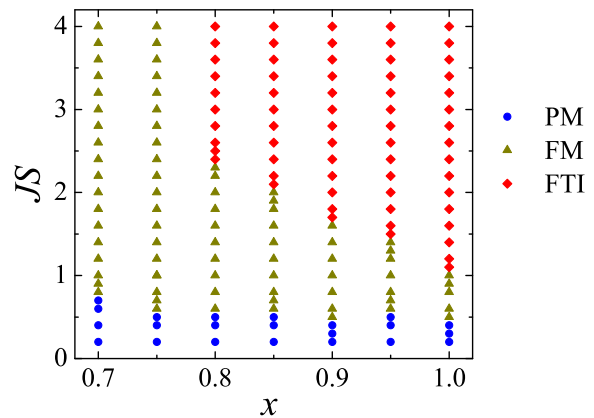}
\caption{(Color online) Phase diagram at electron quarter filling $n=0.5$ ($\lambda=0.5$). Abbreviations PM, FM, FTI denote paramagnetic metal,  ferromagnetic metal, and ferromagnetic topological insulator, respectively. }
\label{fig5}
\end{figure}

In Fig. \ref{fig5} we plot the phase diagram at electron quarter filling $n=0.5$.
The insulating state exists only at strong SE coupling and large values of magnetic impurity doping ($x \gtrsim 0.8$). At small values of $x$, only metallic state
exists. The insulating state is ferromagnetic, since strong SE coupling energetically favors the parallel alignment of electron spins like in the DE mechanism  \cite{Dagotto,Nham1,Yunoki}.
In the ferromagnetic insulator (FI), only the lowest band is fully occupied, and three other bands are empty. It turns out that the FI is topological since the Chern number calculated by Eq. (\ref{chern}) gives $C=1$ for the lowest band. This yields the QAH effect.
First principle calculations for real material Bi$_2$Se$_3$ doped with Cr ions  also reveal the QAH effect  \cite{YuFang,ZhangWang1}.
The phase diagram plotted in Fig. \ref{fig5} also shows that doping of magnetic impurities can drive the topological FI to non-topological ferromagnetic metal (FM). However, this topological phase transition is actually an insulator-metal transition. At the phase boundary, the gap closes. However, a further decrease of magnetic doping does not open the gap again, because electron filling is fixed $n=0.5$ and the chemical potential lies within an energy band.
The phase transition at electron quarter filling is quite different in comparison with the magnetic topological phase transition at electron half filling. At electron half filling the topological invariant remains the same across the magnetic phase transition, whereas at electron quarter filling, the spontaneous ferromagnetic magnetization is maintained across the insulator-metal transition, and the non-trivial topological invariant appears in the insulating side only. Doping of magnetic impurities away from full filling suppresses the gap, hence simultaneously destroys the topological invariant.
The anomalous Hall effect was also suggested to exist in conduction ferromagnets, however it cannot be quantized in metals \cite{Jungwirth1}.

Figure \ref{fig5} also shows a magnetic topological phase transition driven by SE at fixed magnetic doping. When the SE is weak, the ground state is paramagnetic metal (PM) although the SOC is present. Actually, the SOC opens a band gap only at electron half filling. Therefore at quarter filling, the SOC does not affect the metallic properties. Both the metal-insulator and the magnetic transitions are driven solely by the SE. However, the SOC causes non-trivial topological invariants of two lowest bands. One lowest band has the Chern number $C=1$, and the other one has $C=-1$. Since the two lowest bands have opposite spins, the QSH effect occurs at electron half filling. When the two lowest bands are separated by a gap, the ground state
is also insulator at electron quarter filling. Since its topological invariant is integer, the QAH effect occurs. The separation of two lowest bands at electron quarter filling also indicates the fully ferromagnetic state. This can be achieved by strong SE \cite{Tien}. Therefore the QAH effect occurs only at the FI state. However, the SE coupling separates two lowest bands only at dense magnetic doping. At dilute magnetic doping, the SE is valid only at a small number of lattice sites, and in an average manner, it cannot open a band gap at electron quarter filling. In real MTI materials, the QAH effect was observed at certain range of magnetic impurity
concentration \cite{YuFang,ZhangWang1,ChangXue,Jin,ZhangYao}.

\subsection{Equal filling $n=x$}

\begin{figure}[t]
\includegraphics[width=0.48\textwidth]{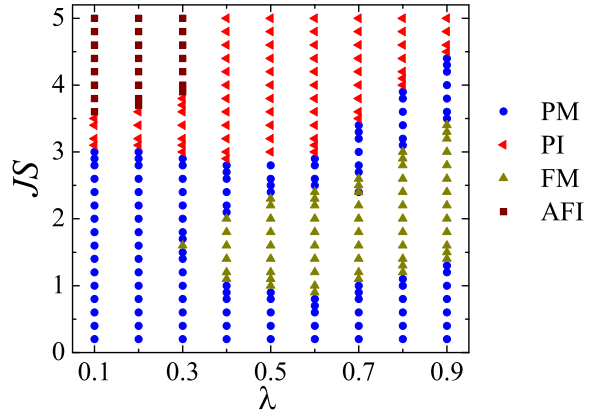}
\caption{(Color online) Phase diagram at equal filling $n=x=0.3$. Abbreviations PM, PI, FM, AFI denote paramagnetic metal, paramagnetic insulator,  ferromagnetic metal, and antiferromagnetic insulator, respectively. All insulating phases are topologically trivial. }
\label{fig6}
\end{figure}

\begin{figure}[t]
\includegraphics[width=0.48\textwidth]{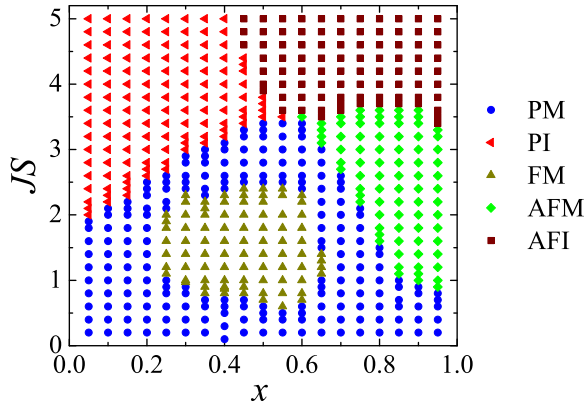}
\caption{(Color online) Phase diagram at equal filling $n=x$ ($\lambda=0.5$). Abbreviations PM, PI, FM, AFM, AFI denote paramagnetic metal, paramagnetic insulator,  ferromagnetic metal, antiferromagnetic metal, and antiferromagnetic insulator, respectively. All insulating phases are topologically trivial.}
\label{fig7}
\end{figure}

In this filling case, the concentration of electrons (holes) is equal to the concentration of magnetic dopants.
The extreme case $n=x=1$ is non-disorder and was previously studied \cite{Tien}.
In Fig. \ref{fig6} we plot the phase diagram at a fixed equal filling $n=x<1$. It exhibits different magnetic states depending on the SOC and the SE.
As we have previously discussed, the SOC opens a band gap only at electron half filling $n=1$ regardless of the SOC strength.
When filling $n=x<1$, the valence band is partially occupied, therefore the ground state is metal. Weak SE does not change this paramagnetic metal (PM). However, the SE polarizes electron spins and shifts the energy bands of opposite spins in opposite directions. This effect of the SE looks like the one of an external magnetic field. Actually, in a mean field approximation, the SE can be treated as a magnetic field.
As a consequence, depending on the relation between the SOC and the SE, the ground state may become FM as can be seen in  Fig. \ref{fig6} (see also Fig. \ref{fig2}). This phase transition is similar to the one obtained in the interplay between the SOC and
external magnetic field \cite{Bo}.
With further increasing SE, a band gap can be opened by the SE, and the ground state becomes paramagnetic insulator (PI). Actually,
Fig. \ref{fig2} also shows when the SE increases, the ferromagnetic state occurs not at a fixed electron filling. It moves toward the domain of lower electron filling as the SE increases. Therefore, when the magnetic doping $x=n$ is fixed, the FM state only occurs in a finite range of the SE.
When the SE is strong enough, the ground state is AFI. Indeed, upon magnetic doping, the AFI at electron half filling is split into two AFIs at fillings $n=x$ and $n=2-x$. In Fig. \ref{fig7} one can also see the impact of magnetic doping on the magnetic states at equal filling $n=x$. The FM state exists only in a finite range of $x$, because the band shift due to the SE lowers the energy band of one spin component, and hence it can maintain the FM state only at certain electron filling $n$. Since $n=x$, as $x$ varies, the electron filling $n$ varies too. Therefore, the phases presented in Fig. \ref{fig7} have varying electron filling, from almost empty filling to almost half filling. The insulating state only exists  when the SE is strong enough.
A strong SE aligns spins of conduction electrons and magnetic moments of impurities. Since the numbers of conduction electrons and of magnetic dopants are the same, there are no free conduction electrons. As a consequence, the insulating state is established.
In the domain of dilute magnetic doping, the insulator is paramagnetic, while in the opposite domain, when the magnetic doping is dense, it is antiferromagnetic. This yields a magnetic phase transition driven by magnetic dopants.
In the case of dense magnetic doping, the ground-state energy is optimized when the AF state is formed like in the limit case $n=x=1$. However, in the dilute doping case, the aligning orientation of electron spins at each lattice site is random. Therefore the macroscopic magnetization vanishes and the PI is established.
We want to emphasize that the magnetic phase transtion driven by magnetic dopants occurs not at a fixed electron filling $n$, but at the constraint $n=x$. In the insulating states at $n=x$, the Chern number calculated by Eq. (\ref{chern}) vanishes.
Although magnetic dopants can maintain the insulating states at equal filling $n=x$, and they can drive the magnetic phase transition from PI to AFI, neither QAH nor QSH effect occurs.  Nevertheless, the phase diagram at equal filling $n=x$ shows rich phase diagrams. Despite the SOC does not cause any topologically nontrivial insulator at $n=x$, its interplay with magnetic dopants gives rise to rich magnetic phases.

\section{Conclusion}

We have studied the impact of magnetic dopants on the magnetic and topological phases which could occur in MTIs.
When magnetic impurities are doped into MTIs, they are coupled with conduction electrons via the SE, and simultaneously introduce disorder and inhomogeneity. The interplay between the random SE and the SOC causes rich magnetic and topological phases in MTIs.
However, non-trivial topology of the insulating ground state exists only at electron half and quarter fillings. At electron half filling the AFI is stable between the PI and AFM, when the SE strength increases. It exhibits the QSH effect that is robust against the magnetic impurity doping. However, disorder and inhomogeneity which are introduced by magnetic dopants induce the AFM at strong SE, while in the non-disordered case, the AFI is instead established. Actually, the AFI at electron half filling is split into two AFIs in the electron and hole domains upon magnetic doping. Although the AFI is topologically nontrivial at electron half filling, its split AFI states upon magnetic doping are topologically trivial. At electron quarter filling, the QAH effect could occur at the strong SE and dense magnetic doping. However, the magnetic doping also drives the FI to the FM, when it decreases, and therefore it completely suppresses the QAH effect at its appropriate value. These findings reveal that magnetic dopants impact differently on the topological properties of the MTIs, depending on electron filling. At electron half filling the topological invariant is robust against magnetic dopants, while at electron quarter filling it is suppressed by magnetic doping.
In addition to the electron half and quarter fillings, we also observed the insulating ground states at equal fillings (i.e.,  the concentration of electrons (holes) is equal to the concentration of magnetic dopants). However, the insulating states are  topologically trivial. In comparison with the non-disordered case, the phase diagram becomes very rich. Disorder and inhomogeneity cause different magnetic orderings in both insulating and metallic states.

Despite the explicit presence of magnetic impurities, the proposed model is also appropriate for intrinsic MTIs, where instead of magnetic impurities, the d-band correlated electrons establish magnetic long-range ordering \cite{LiLi,ZhangShi}.
The intrinsic MTIs were recently discovered and have attracted intensive attention \cite{LiLi,ZhangShi}.
Actually, in the intrinsic MTIs
only the spin degree of freedom of the d-band correlated electrons is relevant for establishing magnetism, and the charge degree of freedom can be discarded. The SE between the d-band correlated electrons and conduction electrons may interplay with the SOC of conduction electrons and this emerges the topologically nontrivial magnetic ground state \cite{Tien}.
We leave this problem for further studies.

\section*{Acknowledgement}

This research is funded by Vietnam National Foundation for Science and Technology Development (NAFOSTED) under Grant No 103.01-2019.309.
In addition, Tuan-Minh, Kim-Thanh and Minh-Tien were supported by the International Center of Physics, Institute of Physics (VAST) at Hanoi
under Grant No. ICP.2019.03, while Thanh-Mai was supported by the International Center for Theoretical Physics at Trieste and the Asian-Pacific Center for Theoretical Physics at Pohang.

\end{document}